\documentclass[11pt]{article}
\usepackage[margin=1in]{geometry}
\usepackage{amsmath,amssymb,amsthm}
\usepackage{graphicx,booktabs,array}
\usepackage[authoryear,round]{natbib}
\usepackage[hidelinks]{hyperref}
\usepackage{listings}
\newtheorem{theorem}{Theorem}
\newtheorem{proposition}{Proposition}
\newtheorem{corollary}{Corollary}
\theoremstyle{remark}
\newtheorem{remark}{Remark}
\newcommand{\PATE}{\mathrm{PATE}}
\title{Generalizing Couple-Randomized Trials Under Limited Overlap and Missing Partner Linkage}
\author{Subir Hait\\[2pt]
\small Department of Counseling, Educational Psychology, and Special Education\\
\small Michigan State University, East Lansing, Michigan 48824, USA\\
\small \texttt{haitsubi@msu.edu} \quad ORCID 0009-0004-9871-9677}
\date{November 2026}

\begin{document}
\maketitle

\begin{abstract}
We study how to generalize couple-randomized trials when some couple types are absent from the trial and the target survey reports partner characteristics without linking spouses. Under an additive effect model with unrestricted partner-profile terms, the target average effect is identified exactly when the wife and husband margins assign equal mass to each connected component of the trial support graph. A specified interaction is identified when its alternating contrast around at least one cycle is nonzero. Without linkage, we obtain sharp bounds by optimizing over pairings consistent with the survey margins. A spousal odds ratio can narrow the bounds for a binary interaction. We derive joint bounds for an index-patient design, quantify bias from common shortcuts, and estimate effects and bounds using Bayesian dyadic multilevel regression and poststratification. In a worked example based on 3,826 Health and Retirement Study couples and simulated trial outcomes, an index-patient trial and an unlinked survey yielded a 2.93 mmHg wide identified set for the couple-average effect. Linking the survey narrowed it to 2.38 mmHg. Adding 300 couples without diabetes narrowed the unlinked set to 0.28 mmHg and yielded point identification with linkage under the fitted model.
\end{abstract}


\section{Introduction}

Many health programs are delivered to couples rather than to individuals \citep{martire2010,ardenclose2017}. Examples include couple-based programs for diet, physical activity, smoking, weight loss, and the self-management of chronic illness such as type 2 diabetes \citep{ardenclose2017,martire2010,trief2016}. Spouses share meals, routines, and homes, and they also resemble each other in health and in health behavior \citep{meyler2007}. For example, a person whose spouse has diabetes has a higher risk of diabetes \citep{leong2014}. Research also shows that when one spouse changes a health behavior, the other spouse often changes too \citep{falba2008,jackson2015}. For this reason, researchers have argued that partners should be part of behavior change programs \citep{jackson2015,martire2010}.

In a couple-randomized trial, the couple is randomized and both partners are measured \citep{trief2016,ardenclose2017}. The effect on one partner can depend on that partner's own traits and also on the traits of the other partner, which is the actor and partner structure of dyadic data \citep{kenny2006,cook2005}. Couple programs can also benefit the partner who is not the patient, for example by lowering the partner's diabetes distress \citep{trief2019}. For example, a diet program may work better for a wife whose husband also has diabetes, because both partners change the shared meals. This is plausible because spouses tend to change their health habits together \citep{falba2008,jackson2015}.

Couple trials rarely enroll a representative set of couples. Recruitment often runs through an index patient, for example a person with poorly controlled diabetes \citep{trief2016}. Because spouses resemble each other \citep{meyler2007,leong2014}, the partner's traits are selected too. The trial average can then differ from the effect in the population of couples that matters for policy \citep{imai2008,stuart2011}. To move the trial result to that population, we need a population survey that describes the target couples \citep{ackerman2021,degtiar2023}. The Health and Retirement Study (HRS) is one such survey for older couples, because it interviews both spouses \citep{sonnega2014,falba2008}. Many other data sources, however, describe each person without linking spouses \citep{ridder2007,dorazio2006}. Also, the trial itself may never enroll some types of couples, which breaks the positivity condition that generalization methods need \citep{petersen2012,degtiar2023}.

The above discussion implies that a couple trial raises three problems at once. Each partner's effect can depend on the other partner's traits \citep{cook2005,kenny2006}. The trial may not enroll every type of couple \citep{trief2016,petersen2012}. The target survey may describe each partner without linking spouses \citep{ridder2007,dorazio2006}. These three problems interact. Partner traits can matter even when the exact pairing does not, and the pairing matters only through actor-by-partner interactions. These interactions are exactly what a selectively recruited trial may be unable to learn.

\section{Literature Review}

Randomized trials give unbiased effects for the people they enroll \citep{rubin1974}, but those people may not represent the population of interest \citep{imai2008,stuart2011}. Previous research shows how to transport a trial effect to a target population by weighting \citep{cole2010,stuart2011,westreich2017}, by subclassification \citep{tipton2013,omuircheartaigh2014}, or by outcome modeling \citep{kern2016,dahabreh2020}. Pearl and Bareinboim gave graphical conditions for when such transport is possible \citep{pearl2014}. Degtiar and Rose reviewed these methods and the assumptions they need \citep{degtiar2023}. Later work extended the methods to targets described by complex surveys \citep{ackerman2021} and to trials that randomize whole clusters \citep{dahabreh2022cluster,robertson2024}. All of these methods need positivity, which means that every type of person in the target must have a chance to be in the trial \citep{petersen2012,westreich2017}. Sensitivity analysis for effect modifiers that are not measured in the target has also been studied \citep{nguyen2017,dahabreh2023,huang2024,asiaee2026}.

Multilevel regression and poststratification (MRP) gives model-based estimates for a target table \citep{gelman1997,park2004}. It works by fitting a multilevel model to cell means and then weighting the cells by their population shares \citep{gelman2007,si2020}. MRP has produced good estimates from very unrepresentative samples, for example from opt-in election polls \citep{wang2015}. It can also estimate deep interactions among small subgroups \citep{ghitza2013}. In health research, MRP has been used to correct for nonparticipation in highly selected samples \citep{downes2018}. Recent work uses MRP to generalize treatment effects from trials \citep{kennedy2021,gao2021te}. Research also shows that MRP can be sensitive to its prior on the variance when the sample is selective \citep{bruch2022,gao2021structured}. Bayesian computation for these models is now routine in Stan \citep{carpenter2017}.

When two data sets cannot be linked, joint distributions are only partially identified \citep{manski2003,ridder2007}. The statistical matching literature studies what can be learned about a joint table from its margins \citep{dorazio2006}. Fan, Sherman, and Shum derived sharp bounds on treatment effects when outcomes and covariates come from separate data sets \citep{fan2014}. Such bounds are often found by optimal transport \citep{galichon2016}, and for two binary variables they reduce to the Fr\'echet--Hoeffding bounds \citep{plackett1965,dorazio2006}.

Dyadic data bring their own problems. Treating one partner can change the other partner's outcome, so causal effects must be defined at the level of the couple \citep{hudgens2008}. The actor--partner interdependence model separates the effect of a person's own traits from the effect of the partner's traits \citep{cook2005,kenny2006}. Spouses are concordant in health \citep{meyler2007,leong2014}, so a trial that recruits through one partner also selects the other partner \citep{trief2016}. To our knowledge, previous research has not studied generalization when partner traits modify the effect, when the trial does not enroll every type of couple, and when the target survey does not link spouses.

\section{Objectives}

The purpose of the current study is fourfold. The first objective is to find what a couple trial identifies when it does not enroll every type of couple. We show that, under partner-additive effects, the target effect is identified when each connected component of the trial's support graph has equal target weight on the wife and husband sides (Theorem~\ref{thm:balance}). An actor-by-partner interaction needs a cycle in that graph (Proposition~\ref{prop:cycle}). The second objective is to find what can be learned when the survey does not link spouses. We show that the target effect has sharp bounds given by a discrete optimal transport problem (Proposition~\ref{prop:ot}). For one binary interaction the bounds have a closed form, and a spousal odds ratio serves as a sensitivity parameter. When both problems occur together, the joint bounds sit at the corners of the two ranges (Corollary~\ref{cor:joint}). The third objective is to measure the bias of three shortcuts: adjusting only for each person's own traits, building the couple table from the margins, and the shrinkage that standard MRP applies when the trial is selected on the modifiers (Propositions~\ref{prop:actor}, \ref{prop:indep}, and~\ref{prop:shrink}). The fourth objective is to estimate the effects and the bounds with Bayesian dyadic MRP fitted in Stan \citep{carpenter2017}.

The implications of these results are practical. The two gaps call for different remedies. Missing couple types call for different recruitment. Missing linkage calls for better population data or an outside estimate of spousal concordance. Section~\ref{sec:example} shows how to tell which gap blocks a given decision.

We study the methods in a plasmode simulation \citep{franklin2014} built on couples from the 2018 Health and Retirement Study (HRS) \citep{sonnega2014,randhrs2026}. The covariates, strata, clusters, and weights are real. The trial outcomes are simulated, so the true population effect is known.

\section{Notation and Assumptions}\label{sec:setup}

A trial sample of couples is randomized to treatment $A=1$ or control $A=0$. The outcome $Y_p$ is measured on each partner $p\in\{w,h\}$ (wife, husband). A target survey of couples is drawn by a complex design with strata, clusters, and weights. Both sources record discrete traits $X=(R,X_w,X_h)$: couple race and ethnicity $R$, the wife's profile $X_w$, and the husband's profile $X_h$. In our application, each profile combines age group (under 65, 65--74, 75 and older), body mass index group (under 25, 25--30, 30 and over), and diagnosed diabetes. So there are $4\times18\times18=1{,}296$ couple cells $j$. We write $a$ for a wife profile, $b$ for a husband profile, and $D_w$, $D_h$ for the two diabetes indicators.

Treatment is assigned to the couple, so each partner's potential outcome $Y_p(a)$ already includes any spillover from the other partner. We assume no interference between couples \citep{hudgens2008}. Let $\tau_p(j)=E\{Y_p(1)-Y_p(0)\mid X=j\}$, and let $\pi_T(j)$ be the share of target couples in cell $j$. The estimands are
\begin{equation}
\PATE_p=\sum_j \pi_T(j)\,\tau_p(j), \qquad \PATE_c=\tfrac12(\PATE_w+\PATE_h).
\end{equation}
We make three assumptions. (A1) Treatment is randomized within the trial. (A2) Dyadic transportability holds: $E\{Y_p(1)-Y_p(0)\mid X=j,\text{trial}\}=E\{Y_p(1)-Y_p(0)\mid X=j,\text{target}\}$, with both partners' traits in $X$. (A3) The survey design is valid for $\pi_T$. Section~\ref{sec:overlap} replaces the usual positivity assumption with a weaker condition, and Section~\ref{sec:sens} studies what happens when (A2) fails.

\section{Identification}\label{sec:ident}

In this section, we first give the bias of two shortcuts. We then give bounds for missing linkage and identification results for limited overlap.

\subsection{Two shortcuts}

\begin{proposition}[Actor-only adjustment]\label{prop:actor}
Write $j=(a,b)$ with $a$ the actor cell $(R,X_p)$ and $b$ the partner's profile. An analysis that adjusts only for the actor cell and poststratifies on the target actor margin converges to $\sum_a\pi_T(a)\sum_b\pi_S(b\mid a)\tau_p(a,b)$, where $\pi_S$ is the trial distribution. Its bias is
\begin{equation}
\sum_a \pi_T(a)\sum_b \tau_p(a,b)\{\pi_S(b\mid a)-\pi_T(b\mid a)\}.
\end{equation}
\end{proposition}
The bias is zero if partner traits do not modify the effect, or if the trial and the target have the same partner mix within each actor cell. Index-patient recruitment breaks the second condition, because spouses are concordant.

\begin{proposition}[Independent pairing]\label{prop:indep}
Let $\pi_{\rm ind}(r,a,b)=\pi_T(r)\pi_T(a\mid r)\pi_T(b\mid r)$. If $\tau_p(r,a,b)=f_p(r,a)+g_p(r,b)+\theta_p D_w D_h$, then
\begin{equation}
\sum_j\pi_{\rm ind}(j)\tau_p(j)-\PATE_p=-\theta_p\sum_r\pi_T(r)\,\mathrm{Cov}_T(D_w,D_h\mid r).
\end{equation}
\end{proposition}
The independent table keeps every one-partner margin, so only the interaction contributes to the bias. In the HRS couples, the weighted within-race covariance of the two diabetes indicators averages $0.0127$. So even $\theta_p=-2$ mmHg moves the estimate by only about $0.03$ mmHg.

\subsection{Missing linkage: sharp bounds}\label{sec:linkage}

Suppose the survey gives the race shares $\pi_T(r)$ and the wife and husband profile margins $p_w(a\mid r)$ and $p_h(b\mid r)$, but not the pairing. Let $\Gamma_r$ be the set of joint tables $\gamma_r(a,b)$ with these margins.

\begin{proposition}[Sharp bounds without linkage]\label{prop:ot}
Under (A1)--(A3) and known cell effects, the identified set for $\PATE_p$ is the interval $[L,U]$ with
\begin{equation}
L=\sum_r\pi_T(r)\min_{\gamma_r\in\Gamma_r}\sum_{a,b}\gamma_r(a,b)\tau_p(r,a,b),\qquad
U=\sum_r\pi_T(r)\max_{\gamma_r\in\Gamma_r}\sum_{a,b}\gamma_r(a,b)\tau_p(r,a,b).
\end{equation}
Each inner problem is a discrete optimal transport problem, and the bounds are sharp.
\end{proposition}

\begin{corollary}[One binary interaction]\label{cor:frechet}
If $\tau_p(r,a,b)=f_p(r,a)+g_p(r,b)+\theta_p D_wD_h$, then $\PATE_p=\Psi_{\rm add}+\theta_p\sum_r\pi_T(r)P_{11}(r)$, where $\Psi_{\rm add}$ depends only on the margins and $P_{11}(r)=P(D_w=1,D_h=1\mid r)$ ranges over the Fr\'echet--Hoeffding interval $[\max(0,q_w+q_h-1),\min(q_w,q_h)]$, with $q_w,q_h$ the diabetes shares in race group $r$. If the spousal odds ratio $\psi$ is known, $P_{11}$ is the root of Plackett's equation \citep{plackett1965},
\begin{equation}
P_{11}(\psi)=\frac{S-\sqrt{S^2-4\psi(\psi-1)q_wq_h}}{2(\psi-1)},\qquad S=1+(\psi-1)(q_w+q_h),
\end{equation}
and $\psi=1$ gives independent pairing.
\end{corollary}

\subsection{Limited overlap: what the trial identifies}\label{sec:overlap}

We work within one race group. Let $S$ be the set of couple types $(a,b)$ that the trial supports in both arms, and let $G$ be the bipartite graph on wife profiles and husband profiles with an edge for each $(a,b)\in S$. Under (A1)--(A2), the trial identifies $\tau(a,b)$ for $(a,b)\in S$.

\begin{theorem}[Overlap balance]\label{thm:balance}
Suppose $\tau(a,b)=\alpha_a+\beta_b$ for all $a,b$. Then\[\Psi=\sum_{a,b}\pi_T(a,b)\tau(a,b)=\sum_a p_w(a)\alpha_a+\sum_b p_h(b)\beta_b\]is identified if and only if every connected component $K$ of $G$ (an isolated profile is its own component) satisfies $\sum_{a\in K}p_w(a)=\sum_{b\in K}p_h(b)$.
\end{theorem}

A connected graph is enough, even when many couple types are missing from the trial. The condition uses only the two margins, so an additive effect can be transported to a survey that does not link spouses. For example, with diabetes alone, a trial that supports types 00, 01, and 11 gives $\tau_{10}=\tau_{00}+\tau_{11}-\tau_{01}$.

\begin{proposition}[Interactions need cycles]\label{prop:cycle}
Let $\tau(a,b)=\alpha_a+\beta_b+\eta\,h(a,b)$ and suppose the balance condition holds. Then $\eta$ is identified if and only if some cycle of $G$ has a nonzero alternating sum of $h$. Otherwise there are $f,g$ with $h(a,b)=f(a)+g(b)$ on $S$, and the identified set is the line
\begin{equation}
\Psi(\eta)=\Psi(0)+\eta\,\Delta,\qquad \Delta=\sum_{a,b}\pi_T(a,b)\{h(a,b)-f(a)-g(b)\},
\end{equation}
where $\Psi(0)=\sum_{a,b}\pi_T(a,b)\{\alpha_a+\beta_b+\eta f(a)+\eta g(b)\}$ is the additive extension of the supported effects. $\Psi(0)$ is determined by the effects on $S$ and does not depend on $\eta$.
\end{proposition}

Now consider an index-patient trial that enrolls only couples with at least one partner with diabetes. On $S$ we have $D_wD_h=D_w+D_h-1$, so $\Delta=\pi_T(00)$, the target share of couples in which neither partner has diabetes. Then $\tau_{00}=\tau_{10}+\tau_{01}-\tau_{11}+\eta$ and $\PATE(\eta)=\PATE(0)+\eta\,\pi_T(00)$, with
\begin{equation}\label{eq:pate0}
\PATE(0)=\sum_j\pi_T(j)\{\tau(j)-\eta(1-D_w)(1-D_h)\}=\PATE(\eta)-\eta\,\pi_T(00).
\end{equation}
This baseline keeps the effects in the three enrolled diabetes types. It replaces the effect in couples without diabetes by its additive extrapolation $\tau_{10}+\tau_{01}-\tau_{11}$. Simply dropping the term $\eta D_wD_h$ from every cell is not the same. That would change the effects of enrolled couples in which both partners have diabetes, and it differs from \eqref{eq:pate0} by $\eta\{\pi_T(00)-\pi_T(11)\}$, about 1.05 mmHg at $\eta=-2$ with the HRS margins.

\begin{remark}[Covariate models]\label{rem:linear}
Theorem~\ref{thm:balance} and Proposition~\ref{prop:cycle} allow an unrestricted effect for every wife and husband profile. Our Stan model instead uses fewer covariate terms, $\tau(j)=z(j)^\top\theta$. In that case a target functional $c^\top\theta$ is identified if and only if $c$ lies in the row space of the trial design, which is the usual condition for linear estimability. A covariate model can then identify the target when the profile-level graph does not, but the result rests on the model's functional form. Section~\ref{sec:example} reports which functionals the fitted model identifies.
\end{remark}

\begin{corollary}[Joint bounds]\label{cor:joint}
For the index-patient case without linkage, $\pi_T(00)=\sum_r\pi_T(r)\{1-q_w-q_h+P_{11}(r)\}$ with $P_{11}(r)$ in its Fr\'echet range, $\PATE(0)$ depends only on the margins, and $\eta$ lies in a stated range $[\eta_-,\eta_+]$. The function $(\eta,\pi_T(00))\mapsto\PATE(0)+\eta\,\pi_T(00)$ is bilinear, so the sharp joint bounds are attained at the four corners of the two ranges.
\end{corollary}

In short, recruitment decides whether the interaction can be learned, and linkage decides how often it applies in the target. Recruiting the missing couple types closes a cycle and identifies $\eta$. Linked records or a concordance estimate pin down $P_{11}$.

\section{Estimation: Bayesian Dyadic MRP}\label{sec:est}

\subsection{Outcome model}
For couple $i$ in cell $j$ and arm $a$, we model the two outcomes as bivariate normal with means $\mu_p(j)+a\,\tau_p(j)$ and covariance $\Sigma$. The covariance carries the residual correlation within the couple. Each partner's effect is $\tau_p(j)=\tau_{0p}+z_p(j)^\top\theta_p$, where $z_p(j)$ holds 14 reference-coded columns: own age (2), own BMI (2), own diabetes, partner age (2), partner BMI (2), partner diabetes, own-by-partner diabetes, and race (3). The prognostic part has terms for own age, own BMI, own diabetes, partner diabetes, and race. By default we partially pool the wife's and husband's coefficients, $\theta_{pk}=\bar\theta_k+\sigma_{b(k)}\varepsilon_{pk}$. The priors are $\bar\theta_k\sim N(0,5^2)$, $\sigma_b\sim N^+(0,1.5^2)$ for each of eight batches, $N(0,10^2)$ for $\tau_{0p}$ and the prognostic coefficients, $N^+(0,20^2)$ for the residual scales, and LKJ(2) for the correlation. Because the covariates are discrete, the likelihood depends only on cell-by-arm means and the pooled scatter matrix. So fitting time does not grow with trial size.

\subsection{Composition model}
Let $c_j$ be the weighted survey count in cell $j$, rescaled to sum to the effective sample size $n_{\rm eff}=n_T/\mathrm{deff}$. Here deff is the median design effect over all factor margins \citep{savitsky2016,williams2021}. We use $\pi_T\mid c\sim\mathrm{Dirichlet}(n_{\rm eff}[(1-\omega)\hat\pi_{\rm HT}+\omega m])$, where $\hat\pi_{\rm HT}$ holds the weighted shares, $m$ is a log-linear base measure with all main effects and two-way interactions, and $\omega=a_0/(a_0+n_{\rm eff})$ with $a_0$ estimated by empirical Bayes. Setting the total concentration to $n_{\rm eff}$ keeps the posterior from claiming more information than the survey holds. The posterior is conjugate, and we draw it in Stan's generated quantities.

\begin{remark}[When the survey's own uncertainty matters]
Treating cell effects as fixed,\[\mathrm{Var}\Big(\sum_j\hat\pi_j\tau(j)\Big)\approx\mathrm{Var}_{\pi_T}\{\tau(X)\}/n_{\rm eff}.\] The survey's share of the posterior variance is small unless effects vary a lot across cells and the survey is small relative to the trial.
\end{remark}

\subsection{Why standard MRP shrinkage is biased under selection}

\begin{proposition}[Shrinkage bias]\label{prop:shrink}
Let $\tilde\tau_j=\lambda_j\hat\tau_j+(1-\lambda_j)\hat\mu$, where $\hat\tau_j$ is unbiased for $\tau_j$, $\lambda_j\in[0,1]$ is fixed ($\lambda_j=0$ for cells without data), and $\hat\mu=\sum_j w_j\hat\tau_j/\sum_j w_j$ with precision weights $w_j$. Then
\begin{equation}
E\Big[\sum_j\pi_T(j)\tilde\tau_j\Big]-\PATE=\sum_j\pi_T(j)(1-\lambda_j)(\bar\mu-\tau_j),\qquad \bar\mu=\frac{\sum_jw_j\tau_j}{\sum_jw_j}.
\end{equation}
\end{proposition}
The weights $w_j$ grow with the number of trial couples in cell $j$. So $\bar\mu$ is close to the trial-average effect, and $\lambda_j$ is smallest in the cells the trial under-samples. If every $\lambda_j$ is near zero, MRP returns the trial average. Standard MRP with zero-mean random effects for each modifier behaves the same way along each batch. The remedy is to avoid a center anchored to the trial, so our model gives each modifier a weakly informative prior instead. Previous research shows that MRP is sensitive to its variance prior under selective samples \citep{bruch2022,gao2021structured}. The proposition makes the direction and the size of this bias explicit for treatment-effect transport.

\subsection{Posterior for bounds}
For each posterior draw of the cell effects, we compute the bounds of Section~\ref{sec:ident} exactly. We use linear programming for Proposition~\ref{prop:ot} and closed forms for Corollaries~\ref{cor:frechet} and~\ref{cor:joint}. For the index-patient design, we compute $\PATE(0)$ in each draw as $\PATE-\eta\,\pi_T(00)$, equation~\eqref{eq:pate0}. This quantity does not change along the one direction that such a trial cannot learn (Section~\ref{sec:example}). We report the posterior means of the two endpoints, the 2.5th percentile of the lower endpoint, and the 97.5th percentile of the upper endpoint.

\section{Sensitivity to an Unmeasured Couple-Level Modifier}\label{sec:sens}

Let a binary couple trait $U$ shift both partners' effects by $\delta$, and let $q_S(j)$ and $q_T(j)$ be its prevalence in cell $j$ in the trial and in the target. Omitting $U$ when transporting the trial's cell effects adds bias $\delta\bar\Delta$ to dyadic MRP, where $\bar\Delta=\sum_j\pi_T(j)\{q_S(j)-q_T(j)\}$. The bias is the same for wives, husbands, and the couple average. This product form also appears in earlier work on generalization \citep{nguyen2017,dahabreh2023,huang2024,asiaee2026}. We report the tipping product $\delta\bar\Delta$ that moves the estimate to a threshold. We benchmark it with an observed trait $V$, using its estimated modification of the couple-average effect and its raw trial--target imbalance. A measured trait has no within-cell imbalance by construction. So its raw imbalance is a reference scale, not an estimate of $\bar\Delta$, and treating an unmeasured trait as that imbalanced within cells is a deliberately strong benchmark.

\section{Simulation Study}\label{sec:sim}

\subsection{Methods}
In this study, we used wave 14 (2018) of the RAND HRS Longitudinal File \citep{randhrs2026}. The target population is community-dwelling, married, different-sex couples in which both spouses were interviewed with positive respondent weights. After we dropped 104 couples with missing covariates, 3,826 couples remained, in 80 strata with 2 sampling error computation units each. The 2018 household weight sums to 24.0 million couples. We used it only to build the pseudo-population, because it is not adjusted for our complete-case restrictions. Kish's design effect from unequal weights is 1.73. Diabetes is reported by 21.1\% of wives and 26.7\% of husbands, and by both partners in 7.3\% of couples. In 59.6\% of couples, neither partner has diabetes.

We copied each couple in proportion to its weight to form a pseudo-population of about 150,000 couples, and we split each real cluster into four. The outcome is the six-month change in systolic blood pressure. The wife's true effect is $-4-2D_w-1.5\,\mathrm{Ob}_w+1.5\,\mathrm{Old}_w-2D_h-2D_wD_h-1.0\,\mathrm{Black}-0.5\,\mathrm{Hisp}+1.0\,\mathrm{Old}_w\mathrm{Ob}_w+\nu$. The husband's effect is defined the same way from a base of $-3$. The last term is not in the fitted model. The population effects are $-5.49$ (wives), $-4.35$ (husbands), and $-4.92$ mmHg (couple average). Each replicate draws a survey and a trial. The survey takes two of eight clusters per stratum and oversamples Black and Hispanic couples two to one. The trial over-enrolls couples with diabetes, and the trial average is about $-7.7$ mmHg. We ran three scenarios. S1 has 1,000 trial couples and about 3,800 survey couples (150 replicates). S2 has 3,000 trial couples and 800 survey couples (150 replicates). S3 is S1 with an unmeasured couple trait, $\delta=-3$ (100 replicates).

\subsection{Data Analysis}
We compared four estimators: the trial difference in means, survey-weighted inverse-odds weighting \citep[IPW;][]{ackerman2021} with a cluster bootstrap, actor-only regression with poststratification, and dyadic MRP. For each estimator we computed bias, root mean squared error (RMSE), the coverage of 95\% intervals, and interval width.

\subsection{Results}

\begin{figure}[t]
\centering
\includegraphics[width=\linewidth]{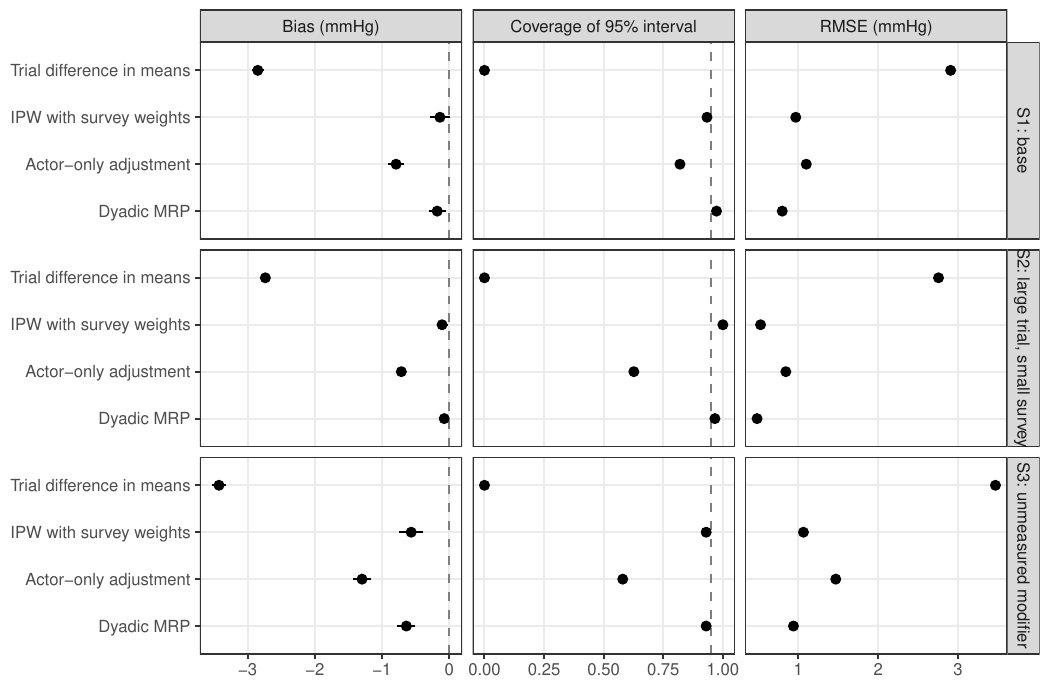}
\caption{Simulation results for the couple-average effect. Bias whiskers are 95\% Monte Carlo intervals; dashed lines mark zero bias and 95\% coverage.}
\label{fig:sim}
\end{figure}

Dyadic MRP removed nearly all of the gap between the trial and the target (Figure~\ref{fig:sim}). In S1 its bias was $-0.18$ mmHg with 97.3\% coverage, against $-2.85$ for the trial difference in means. Its RMSE was 17\% lower than that of IPW (0.80 versus 0.97), and its intervals were 15\% narrower (3.28 versus 3.85 mmHg). Actor-only adjustment left $-0.79$ mmHg, about a quarter of the gap, with 82\% coverage. Its coverage fell to 63\% in S2. In S2 the small survey made the bootstrap intervals of IPW very wide (5.44 versus 1.94 mmHg). Independent pairing and fixed survey shares changed the dyadic MRP estimate by less than 0.03 mmHg. The survey's share of posterior variance averaged 0.65\% in S1 and 3.6\% in S2.

In S3, the quantity in the bias formula is the target-weighted within-cell imbalance, $\bar\Delta=0.243$. So the formula predicts that omitting $U$ contributes $-3\times0.243=-0.73$ mmHg of bias. The raw gap is close (64.6\% of trial couples against 40.1\% of the population, a difference of 0.245) only because $U$ was generated independently of the measured traits. The total bias of dyadic MRP was $-0.64$ mmHg (Monte Carlo SE 0.07), which also includes the model's own error. In a paired check, we refitted the 100 S3 replicates with the effect of $U$ removed from the outcomes. Omitting $U$ contributed $-0.71$ mmHg (Monte Carlo SE 0.01). Ten of 400 fits had any divergent transitions, and the largest $\hat R$ was 1.016.

\paragraph{Shrinkage.} On the same replicates, standard MRP with zero-mean random effects for every modifier level left $-0.72$ mmHg in S1 (88.0\% coverage) and $-0.39$ in the first 100 S2 replicates (89.0\%). Our model left $-0.18$ and $-0.08$ with the same interval width. Saturated cell-level MRP with empirical Bayes shrinkage collapsed toward the trial average ($-2.82$ and $-2.16$ mmHg), and Proposition~\ref{prop:shrink} predicted $-2.80$ and $-2.10$. To isolate pooling, we refitted our model on the first 60 S1 replicates with separate, partial, and complete pooling across partners. We did this under the main truth and under an asymmetric truth in which wives respond to their own diabetes and husbands respond to their wife's diabetes. Couple-average biases ranged from $-0.23$ to $-0.37$ mmHg, all within Monte Carlo error of each other. So the gain over standard MRP comes from not shrinking the modifiers toward zero, not from pooling across partners.

\begin{figure}[t]
\centering
\includegraphics[width=\linewidth]{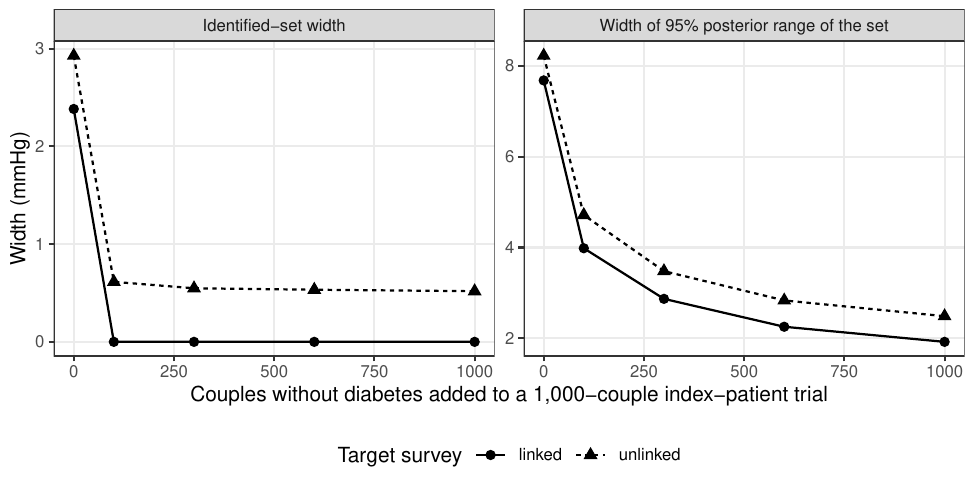}
\caption{Recruitment simulation (40 replicates). Left: posterior mean width of the identified set. Right: width from the 2.5th percentile of the lower endpoint to the 97.5th percentile of the upper endpoint. Linked: the survey links spouses; unlinked: only the partners' margins are known.}
\label{fig:recruit}
\end{figure}

\paragraph{Recruitment.} To see how many couples are enough, we added 0, 100, 300, 600, or 1,000 randomized couples without diabetes to a 1,000-couple index-patient trial. We fitted dyadic MRP to each combination (40 replicates; Figure~\ref{fig:recruit}). We report the posterior mean width of the identified set. We also report the width from the 2.5th percentile of its lower endpoint to the 97.5th percentile of its upper endpoint. With no added couples, the identified set was 2.38 mmHg wide with linkage and 2.93 without, and the posterior range was 7.7 and 8.2 mmHg. Adding 100 couples made the effect a point with linkage and cut the unlinked set to 0.61 mmHg. The posterior ranges fell to 4.0 and 4.7 mmHg. More couples bought precision but not identification. With 1,000 added couples the ranges were 1.9 and 2.5 mmHg, while the unlinked set stayed near 0.5 mmHg. This happens because its width depends on the unknown pairing and not on trial size. The posterior range covered the true effect in 92.5\% to 100\% of replicates at each size, and its midpoint was within 0.25 mmHg of the truth on average. Ten of 200 fits had any divergent transitions.

\section{Worked Example: A Couple-Based Blood Pressure Program}\label{sec:example}

In this example, we take the real 2018 HRS couple file as the target and one simulated trial of 1,000 couples drawn by the S1 recruitment rule. The trial is far from the target. In the trial, 61.6\% of wives and 59.8\% of husbands have diabetes, against 21.1\% and 26.7\% in the target. Also, 42.3\% of trial couples have two partners with diabetes, against 7.3\% in the target. Only 30.0\% of target weight sits in cells with at least three trial couples per arm, and 17.9\% sits in cells with no trial couples. So most of the estimate is carried by the model's structure.

\begin{figure}[t]
\centering
\includegraphics[width=0.85\linewidth]{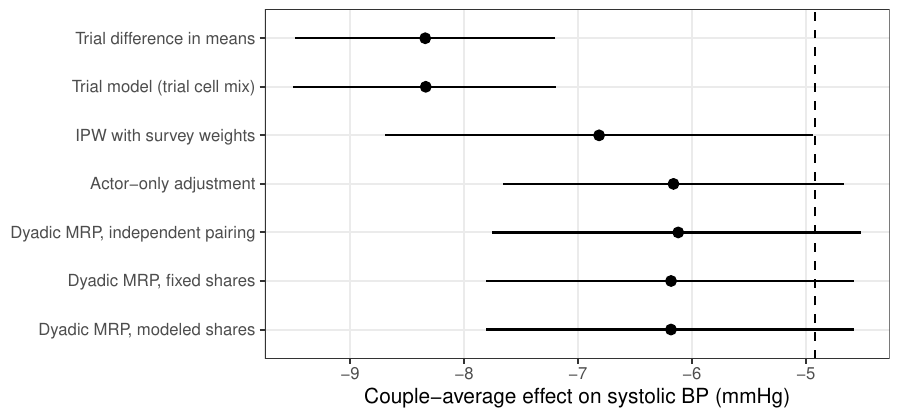}
\caption{Worked example: couple-average effect by estimator with 95\% intervals. The dashed line is the known population effect ($-4.92$ mmHg).}
\label{fig:example}
\end{figure}

Dyadic MRP moved the trial effect of $-8.34$ mmHg to $-6.18$ ($-7.81$ to $-4.58$; Figure~\ref{fig:example}). The true effect is $-4.92$. The fit had no divergent transitions, $\hat R\le1.003$, and bulk effective sample sizes above 1,280. For couples in which neither partner has diabetes, 59.6\% of the target, the estimate was $-6.1$ mmHg against a truth of $-3.7$. This is an extrapolation from thin support. Across the S1 replicates, the interval for that subgroup covered the truth 98\% of the time. An unmeasured couple trait would need about 5.6 times the product for the wife's diabetes ($\delta_V=-1.40$ mmHg, $\Delta_V=0.405$) to shrink the effect to $-3$ mmHg.

\begin{figure}[t]
\centering
\includegraphics[width=0.72\linewidth]{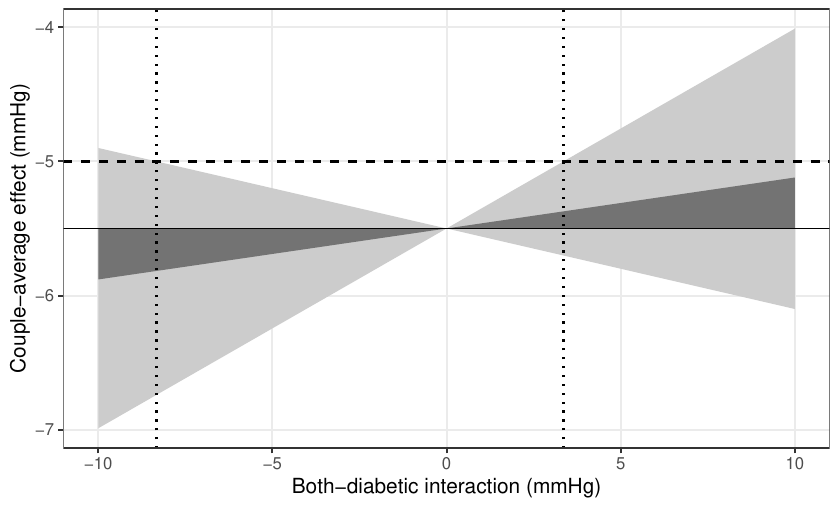}
\caption{When missing linkage changes a decision. Light band: sharp bounds without linkage; dark band: bounds with a spousal odds ratio between 1 and 3. Solid line: independent-pairing estimate ($-5.5$); dashed line: 5 mmHg threshold.}
\label{fig:decision}
\end{figure}

\paragraph{Without linkage.} We then hid the HRS linkage and kept only the margins. The couple-average effect lies between $-6.79$ and $-5.78$ mmHg, a width of 1.01 (Proposition~\ref{prop:ot}). The linear program matched Corollary~\ref{cor:frechet} to within $3\times10^{-12}$ in every draw. The width equals $|\theta|\sum_r\pi_T(r)\min(q_w,q_h)=0.209|\theta|$, so it is driven by the diabetes-by-diabetes coefficient. If we hold an independent-pairing estimate at $-5.5$ mmHg, missing linkage changes a 5 mmHg decision only if that interaction is above $+3.36$ or below $-8.33$ mmHg (Figure~\ref{fig:decision}). A spousal odds ratio between 1 and 3 moves the first threshold to $+13.15$.

\paragraph{Limited overlap and missing linkage together.} We drew a second trial of 1,000 couples only from couples with at least one partner with diabetes. Here identification rests on the fitted covariate model (Remark~\ref{rem:linear}). Each partner's regression design (intercept, 9 prognostic terms, treatment, and 14 treatment-by-modifier terms) has rank 24 of 25 in this trial. The missing direction raises the interaction coefficient and the treatment intercept by $c$ and lowers both diabetes coefficients by $c$. So $\PATE(0)$ in \eqref{eq:pate0} is estimable, but $\PATE$ is not. With 300 added couples without diabetes, the rank is 25 and both are estimable. The profile-level support graph, which uses no covariate model, has 2 to 31 components per race group and is unbalanced with or without the added couples. So Theorem~\ref{thm:balance} alone would not identify either quantity. Because the fitted effects are additive apart from $\eta$, $\PATE(0)$ was the same with the linked and the independent-pairing table.

\begin{figure}[t]
\centering
\includegraphics[width=\linewidth]{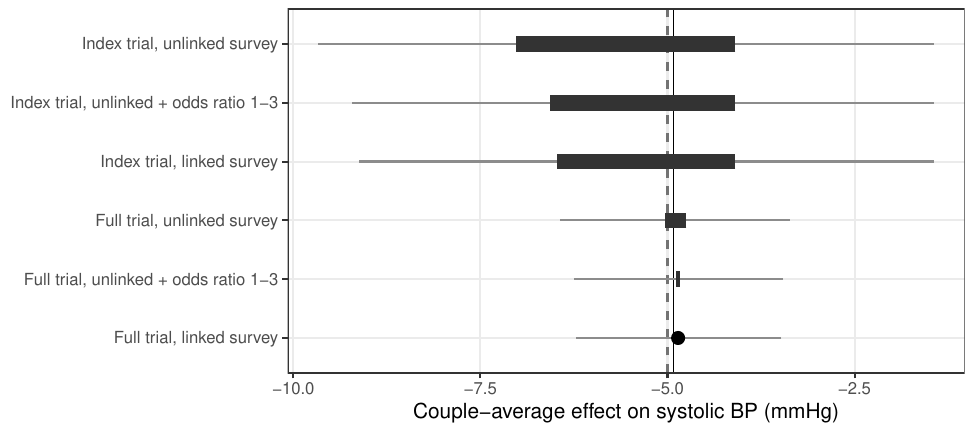}
\caption{What each data source buys. Thick bars: identified sets (posterior means of the endpoints); thin lines: 2.5th percentile of the lower endpoint to 97.5th percentile of the upper endpoint. Solid line: truth; dashed line: 5 mmHg threshold.}
\label{fig:joint}
\end{figure}

The index trial carries no information on $\eta$, so we placed it in $[-4,0]$ mmHg. The share $\pi_T(00)$ is 0.596 with linkage, 0.523 to 0.732 without it, and 0.583 to 0.621 for a spousal odds ratio between 1 and 3. Adding 300 randomized couples without diabetes restores full rank in the fitted model and so identifies $\eta$ (posterior mean $-0.95$, SD 1.38; truth $-2$). Figure~\ref{fig:joint} shows what each data source buys. With the index trial, the identified set is 2.93 mmHg wide without linkage and 2.38 with full linkage, and it straddles the 5 mmHg threshold either way. With the added couples, it is 0.28 mmHg wide without linkage, 0.05 with a concordance range, and, under the fitted model, a point with linkage.

\section{Discussion}

We have seen that the two obstacles in couple trials behave differently. Missing linkage is often less harmful than it looks. Additive effects need only the margins, and the pairing enters only through actor-by-partner interactions, which spousal concordance limits. Missing couple types are often more harmful. An interaction that the trial cannot learn multiplies the target share of the missing type, which was 60\% in our example. In that example, adding 300 couples narrowed the identified set tenfold, while linkage alone narrowed it by a fifth. When a trial treats couples, its results should be generalized to couples, and the analyst should first ask which of the two gaps is present.

We have also seen two lessons about estimation. First, standard MRP shrinkage pulls under-sampled cells toward the trial average when the trial is selected on the modifiers, and Proposition~\ref{prop:shrink} gives the size of that bias. Weakly regularized modifiers removed most of it, and pooling the modifiers across partners mattered little. Second, the parametric model does much of the identifying work. In the worked-example trial, the profile-level support graph had 5 to 34 components per race group and was unbalanced, so the nonparametric partner-additive model would not identify the target effect. Analysts should report trial support and the rank of the fitted design alongside their estimates.

We acknowledge that the study has several limitations. The outcomes are simulated, the HRS weights were not adjusted for our complete-case restrictions, and the interaction range for $\eta$ is illustrative. The bounds assume that each person's couple race group is known. We studied different-sex married couples, and same-sex couples need exchangeable models. Proposition~\ref{prop:shrink} treats the shrinkage weights as fixed.

We hope the findings of this study can inform researchers who design and analyze couple-based trials. The next step is to apply the method to a real couple-randomized trial. The method needs only the trial's cell-by-arm summaries and a survey with both partners' traits, and the HRS is a natural target for trials of older couples. Future research should also study recruitment designs that close the missing cycles at the lowest cost.

\section*{Statements}
\textbf{Data and code.} The RAND HRS Longitudinal File is available from the HRS website after registration. The R and Stan scripts that reproduce every table and figure are included as ancillary files with this preprint and will be posted at \url{https://github.com/subirhait}. The HRS microdata are not redistributed.
\textbf{Funding and conflicts.} No specific funding. The author declares no conflicts of interest.

\bibliographystyle{plainnat}
\bibliography{refs}

@article{ackerman2021,
  author = {Ackerman, Benjamin and Lesko, Catherine R. and Siddique, Juned and Susukida, Ryoko and Stuart, Elizabeth A.},
  title = {Generalizing Randomized Trial Findings to a Target Population Using Complex Survey Population Data},
  journal = {Statistics in Medicine}, year = {2021}, volume = {40}, number = {5}, pages = {1101--1120}}

@article{ardenclose2017,
  author = {Arden-Close, Emily and McGrath, Nuala},
  title = {Health Behaviour Change Interventions for Couples: A Systematic Review},
  journal = {British Journal of Health Psychology}, year = {2017}, volume = {22}, number = {2}, pages = {215--237}}

@misc{asiaee2026,
  author = {Asiaee, Amir and Pal, Samhita and Huling, Jared D.},
  title = {Omitted-Variable Sensitivity Analysis for Generalizing Randomized Trials},
  year = {2026}, howpublished = {arXiv:2603.27788}}

@article{bruch2022,
  author = {Bruch, Christian and Felderer, Barbara},
  title = {Prior Choice for the Variance Parameter in the Multilevel Regression and Poststratification Approach for Highly Selective Data: A {Monte Carlo} Simulation Study},
  journal = {Austrian Journal of Statistics}, year = {2022}, volume = {51}, number = {4}, pages = {76--95}}

@article{carpenter2017,
  author = {Carpenter, Bob and Gelman, Andrew and Hoffman, Matthew D. and Lee, Daniel and Goodrich, Ben and Betancourt, Michael and Brubaker, Marcus and Guo, Jiqiang and Li, Peter and Riddell, Allen},
  title = {{Stan}: A Probabilistic Programming Language},
  journal = {Journal of Statistical Software}, year = {2017}, volume = {76}, number = {1}, pages = {1--32}}

@article{cole2010,
  author = {Cole, Stephen R. and Stuart, Elizabeth A.},
  title = {Generalizing Evidence from Randomized Clinical Trials to Target Populations: The {ACTG} 320 Trial},
  journal = {American Journal of Epidemiology}, year = {2010}, volume = {172}, number = {1}, pages = {107--115}}

@article{dahabreh2020,
  author = {Dahabreh, Issa J. and Robertson, Sarah E. and Steingrimsson, Jon A. and Stuart, Elizabeth A. and Hern{\'a}n, Miguel A.},
  title = {Extending Inferences from a Randomized Trial to a New Target Population},
  journal = {Statistics in Medicine}, year = {2020}, volume = {39}, number = {14}, pages = {1999--2014}}

@misc{dahabreh2022cluster,
  author = {Dahabreh, Issa J. and Robertson, Sarah E. and Steingrimsson, Jon A. and Gravenstein, Stefan and Joyce, Nina R.},
  title = {Extending Inferences from a Cluster Randomized Trial to a Target Population},
  year = {2022}, howpublished = {arXiv:2203.14761}}

@article{dahabreh2023,
  author = {Dahabreh, Issa J. and Robins, James M. and Haneuse, Sebastien J.-P. A. and Saeed, Iman and Robertson, Sarah E. and Stuart, Elizabeth A. and Hern{\'a}n, Miguel A.},
  title = {Sensitivity Analysis Using Bias Functions for Studies Extending Inferences from a Randomized Trial to a Target Population},
  journal = {Statistics in Medicine}, year = {2023}, note = {PMID 36847107}}

@article{franklin2014,
  author = {Franklin, Jessica M. and Schneeweiss, Sebastian and Polinski, Jennifer M. and Rassen, Jeremy A.},
  title = {Plasmode Simulation for the Evaluation of Pharmacoepidemiologic Methods in Complex Healthcare Databases},
  journal = {Computational Statistics \& Data Analysis}, year = {2014}, volume = {72}, pages = {219--226}}

@misc{gao2021te,
  author = {Gao, Yuxiang and Kennedy, Lauren and Simpson, Daniel},
  title = {Treatment Effect Estimation with Multilevel Regression and Poststratification},
  year = {2021}, howpublished = {arXiv:2102.10003}}

@article{gao2021structured,
  author = {Gao, Yuxiang and Kennedy, Lauren and Simpson, Daniel and Gelman, Andrew},
  title = {Improving Multilevel Regression and Poststratification with Structured Priors},
  journal = {Bayesian Analysis}, year = {2021}, volume = {16}, number = {3}, pages = {719--744}}

@article{gelman1997,
  author = {Gelman, Andrew and Little, Thomas C.},
  title = {Poststratification into Many Categories Using Hierarchical Logistic Regression},
  journal = {Survey Methodology}, year = {1997}, volume = {23}, number = {2}, pages = {127--135}}

@article{huang2024,
  author = {Huang, Melody Y.},
  title = {Sensitivity Analysis for the Generalization of Experimental Results},
  journal = {Journal of the Royal Statistical Society Series A}, year = {2024}, volume = {187}, number = {4}, pages = {900--918}}

@article{hudgens2008,
  author = {Hudgens, Michael G. and Halloran, M. Elizabeth},
  title = {Toward Causal Inference with Interference},
  journal = {Journal of the American Statistical Association}, year = {2008}, volume = {103}, number = {482}, pages = {832--842}}

@article{kennedy2021,
  author = {Kennedy, Lauren and Gelman, Andrew},
  title = {Know Your Population and Know Your Model: Using Model-Based Regression and Poststratification to Generalize Findings beyond the Observed Sample},
  journal = {Psychological Methods}, year = {2021}, volume = {26}, number = {5}, pages = {547--558}}

@article{leong2014,
  author = {Leong, Aaron and Rahme, Elham and Dasgupta, Kaberi},
  title = {Spousal Diabetes as a Diabetes Risk Factor: A Systematic Review and Meta-Analysis},
  journal = {BMC Medicine}, year = {2014}, volume = {12}, pages = {12}}

@article{nguyen2017,
  author = {Nguyen, Trang Quynh and Ebnesajjad, Cyrus and Cole, Stephen R. and Stuart, Elizabeth A.},
  title = {Sensitivity Analysis for an Unobserved Moderator in {RCT}-to-Target-Population Generalization of Treatment Effects},
  journal = {Annals of Applied Statistics}, year = {2017}, volume = {11}, number = {1}, pages = {225--247}}

@article{park2004,
  author = {Park, David K. and Gelman, Andrew and Bafumi, Joseph},
  title = {Bayesian Multilevel Estimation with Poststratification: State-Level Estimates from National Polls},
  journal = {Political Analysis}, year = {2004}, volume = {12}, number = {4}, pages = {375--385}}

@article{plackett1965,
  author = {Plackett, R. L.},
  title = {A Class of Bivariate Distributions},
  journal = {Journal of the American Statistical Association}, year = {1965}, volume = {60}, number = {310}, pages = {516--522}}

@manual{randhrs2026,
  title = {{RAND HRS} Longitudinal File 2022 ({V1})},
  author = {{RAND Center for the Study of Aging}}, address = {Santa Monica, CA}, year = {2026},
  note = {Produced with funding from the National Institute on Aging and the Social Security Administration}}

@article{robertson2024,
  author = {Robertson, Sarah E. and Steingrimsson, Jon A. and Dahabreh, Issa J.},
  title = {Cluster Randomized Trials Designed to Support Generalizable Inferences},
  journal = {Evaluation Review}, year = {2024}, volume = {48}, number = {6}, pages = {1088--1114}}

@article{savitsky2016,
  author = {Savitsky, Terrance D. and Toth, Daniell},
  title = {Bayesian Estimation under Informative Sampling},
  journal = {Electronic Journal of Statistics}, year = {2016}, volume = {10}, number = {1}, pages = {1677--1708}}

@article{si2020,
  author = {Si, Yajuan and Trangucci, Rob and Gabry, Jonah Sol and Gelman, Andrew},
  title = {Bayesian Hierarchical Weighting Adjustment and Survey Inference},
  journal = {Survey Methodology}, year = {2020}, volume = {46}, number = {2}, note = {Statistics Canada Catalogue No. 12-001-X}}

@article{sonnega2014,
  author = {Sonnega, Amanda and Faul, Jessica D. and Ofstedal, Mary Beth and Langa, Kenneth M. and Phillips, John W. R. and Weir, David R.},
  title = {Cohort Profile: The {Health and Retirement Study (HRS)}},
  journal = {International Journal of Epidemiology}, year = {2014}, volume = {43}, number = {2}, pages = {576--585}}

@article{stuart2011,
  author = {Stuart, Elizabeth A. and Cole, Stephen R. and Bradshaw, Catherine P. and Leaf, Philip J.},
  title = {The Use of Propensity Scores to Assess the Generalizability of Results from Randomized Trials},
  journal = {Journal of the Royal Statistical Society Series A}, year = {2011}, volume = {174}, number = {2}, pages = {369--386}}

@article{williams2021,
  author = {Williams, Matthew R. and Savitsky, Terrance D.},
  title = {Uncertainty Estimation for Pseudo-{Bayesian} Inference under Complex Sampling},
  journal = {International Statistical Review}, year = {2021}, volume = {89}, number = {1}, pages = {72--107}}

@article{martire2010,
  author = {Martire, Lynn M. and Schulz, Richard and Helgeson, Vicki S. and Small, Brent J. and Saghafi, Ester M.},
  title = {Review and Meta-Analysis of Couple-Oriented Interventions for Chronic Illness},
  journal = {Annals of Behavioral Medicine}, year = {2010}, volume = {40}, number = {3}, pages = {325--342}}

@article{trief2016,
  author = {Trief, Paula M. and Fisher, Lawrence and Sandberg, Jonathan and Cibula, Donald A. and Dimmock, Jacqueline and Hessler, Danielle M. and Forken, Patricia and Weinstock, Ruth S.},
  title = {Health and Psychosocial Outcomes of a Telephonic Couples Behavior Change Intervention in Patients With Poorly Controlled Type 2 Diabetes: A Randomized Clinical Trial},
  journal = {Diabetes Care}, year = {2016}, volume = {39}, number = {12}, pages = {2165--2173}}

@article{trief2019,
  author = {Trief, Paula M. and Fisher, Lawrence and Sandberg, Jonathan and Hessler, Danielle M. and Cibula, Donald A. and Weinstock, Ruth S.},
  title = {Two for One? Effects of a Couples Intervention on Partners of Persons With Type 2 Diabetes: A Randomized Controlled Trial},
  journal = {Diabetic Medicine}, year = {2019}, volume = {36}, number = {4}, pages = {473--481}}

@article{meyler2007,
  author = {Meyler, Deanna and Stimpson, Jim P. and Peek, M. Kristen},
  title = {Health Concordance Within Couples: A Systematic Review},
  journal = {Social Science \& Medicine}, year = {2007}, volume = {64}, number = {11}, pages = {2297--2310}}

@article{falba2008,
  author = {Falba, Tracy A. and Sindelar, Jody L.},
  title = {Spousal Concordance in Health Behavior Change},
  journal = {Health Services Research}, year = {2008}, volume = {43}, number = {1}, pages = {96--116}}

@article{jackson2015,
  author = {Jackson, Sarah E. and Steptoe, Andrew and Wardle, Jane},
  title = {The Influence of Partner's Behavior on Health Behavior Change: The {English} Longitudinal Study of Ageing},
  journal = {JAMA Internal Medicine}, year = {2015}, volume = {175}, number = {3}, pages = {385--392}}

@book{kenny2006,
  author = {Kenny, David A. and Kashy, Deborah A. and Cook, William L.},
  title = {Dyadic Data Analysis}, publisher = {Guilford Press}, address = {New York}, year = {2006}}

@article{cook2005,
  author = {Cook, William L. and Kenny, David A.},
  title = {The Actor--Partner Interdependence Model: A Model of Bidirectional Effects in Developmental Studies},
  journal = {International Journal of Behavioral Development}, year = {2005}, volume = {29}, number = {2}, pages = {101--109}}

@article{westreich2017,
  author = {Westreich, Daniel and Edwards, Jessie K. and Lesko, Catherine R. and Stuart, Elizabeth A. and Cole, Stephen R.},
  title = {Transportability of Trial Results Using Inverse Odds of Sampling Weights},
  journal = {American Journal of Epidemiology}, year = {2017}, volume = {186}, number = {8}, pages = {1010--1014}}

@article{degtiar2023,
  author = {Degtiar, Irina and Rose, Sherri},
  title = {A Review of Generalizability and Transportability},
  journal = {Annual Review of Statistics and Its Application}, year = {2023}, volume = {10}, pages = {501--524}}

@article{kern2016,
  author = {Kern, Holger L. and Stuart, Elizabeth A. and Hill, Jennifer and Green, Donald P.},
  title = {Assessing Methods for Generalizing Experimental Impact Estimates to Target Populations},
  journal = {Journal of Research on Educational Effectiveness}, year = {2016}, volume = {9}, number = {1}, pages = {103--127}}

@article{tipton2013,
  author = {Tipton, Elizabeth},
  title = {Improving Generalizations From Experiments Using Propensity Score Subclassification: Assumptions, Properties, and Contexts},
  journal = {Journal of Educational and Behavioral Statistics}, year = {2013}, volume = {38}, number = {3}, pages = {239--266}}

@article{omuircheartaigh2014,
  author = {O'Muircheartaigh, Colm and Hedges, Larry V.},
  title = {Generalizing From Unrepresentative Experiments: A Stratified Propensity Score Approach},
  journal = {Journal of the Royal Statistical Society: Series C (Applied Statistics)}, year = {2014}, volume = {63}, number = {2}, pages = {195--210}}

@article{imai2008,
  author = {Imai, Kosuke and King, Gary and Stuart, Elizabeth A.},
  title = {Misunderstandings Between Experimentalists and Observationalists About Causal Inference},
  journal = {Journal of the Royal Statistical Society: Series A (Statistics in Society)}, year = {2008}, volume = {171}, number = {2}, pages = {481--502}}

@article{pearl2014,
  author = {Pearl, Judea and Bareinboim, Elias},
  title = {External Validity: From Do-Calculus to Transportability Across Populations},
  journal = {Statistical Science}, year = {2014}, volume = {29}, number = {4}, pages = {579--595}}

@article{rubin1974,
  author = {Rubin, Donald B.},
  title = {Estimating Causal Effects of Treatments in Randomized and Nonrandomized Studies},
  journal = {Journal of Educational Psychology}, year = {1974}, volume = {66}, number = {5}, pages = {688--701}}

@article{petersen2012,
  author = {Petersen, Maya L. and Porter, Kristin E. and Gruber, Susan and Wang, Yue and van der Laan, Mark J.},
  title = {Diagnosing and Responding to Violations in the Positivity Assumption},
  journal = {Statistical Methods in Medical Research}, year = {2012}, volume = {21}, number = {1}, pages = {31--54}}

@article{gelman2007,
  author = {Gelman, Andrew},
  title = {Struggles With Survey Weighting and Regression Modeling},
  journal = {Statistical Science}, year = {2007}, volume = {22}, number = {2}, pages = {153--164}}

@article{ghitza2013,
  author = {Ghitza, Yair and Gelman, Andrew},
  title = {Deep Interactions With {MRP}: Election Turnout and Voting Patterns Among Small Electoral Subgroups},
  journal = {American Journal of Political Science}, year = {2013}, volume = {57}, number = {3}, pages = {762--776}}

@article{wang2015,
  author = {Wang, Wei and Rothschild, David and Goel, Sharad and Gelman, Andrew},
  title = {Forecasting Elections With Non-Representative Polls},
  journal = {International Journal of Forecasting}, year = {2015}, volume = {31}, number = {3}, pages = {980--991}}

@article{downes2018,
  author = {Downes, Marnie and Gurrin, Lyle C. and English, Dallas R. and Pirkis, Jane and Currier, Dianne and Spittal, Matthew J. and Carlin, John B.},
  title = {Multilevel Regression and Poststratification: A Modeling Approach to Estimating Population Quantities From Highly Selected Survey Samples},
  journal = {American Journal of Epidemiology}, year = {2018}, volume = {187}, number = {8}, pages = {1780--1790}}

@book{manski2003,
  author = {Manski, Charles F.},
  title = {Partial Identification of Probability Distributions}, publisher = {Springer}, address = {New York}, year = {2003}}

@incollection{ridder2007,
  author = {Ridder, Geert and Moffitt, Robert},
  title = {The Econometrics of Data Combination},
  booktitle = {Handbook of Econometrics}, editor = {Heckman, James J. and Leamer, Edward E.},
  volume = {6B}, pages = {5469--5547}, publisher = {Elsevier}, address = {Amsterdam}, year = {2007}}

@article{fan2014,
  author = {Fan, Yanqin and Sherman, Robert and Shum, Matthew},
  title = {Identifying Treatment Effects Under Data Combination},
  journal = {Econometrica}, year = {2014}, volume = {82}, number = {2}, pages = {811--822}}

@book{dorazio2006,
  author = {D'Orazio, Marcello and Di Zio, Marco and Scanu, Mauro},
  title = {Statistical Matching: Theory and Practice}, publisher = {Wiley}, address = {Chichester}, year = {2006}}

@book{galichon2016,
  author = {Galichon, Alfred},
  title = {Optimal Transport Methods in Economics}, publisher = {Princeton University Press}, address = {Princeton, NJ}, year = {2016}}

\appendix
\section{Proofs}

\paragraph{Proposition~\ref{prop:actor}.} The actor-only model estimates $E_S\{\tau_p\mid a\}=\sum_b\pi_S(b\mid a)\tau_p(a,b)$ in each actor cell; poststratifying on $\pi_T(a)$ and subtracting $\PATE_p=\sum_a\pi_T(a)\sum_b\pi_T(b\mid a)\tau_p(a,b)$ gives the result. \qed

\paragraph{Proposition~\ref{prop:indep}.} $\pi_{\rm ind}$ and $\pi_T$ share the margins $\pi_T(r,a)$ and $\pi_T(r,b)$, so the additive terms have equal averages. The interaction term averages to $\theta_p\sum_r\pi_T(r)q_wq_h$ under $\pi_{\rm ind}$ and $\theta_p\sum_r\pi_T(r)P_{11}(r)$ under $\pi_T$; their difference is $-\theta_p\sum_r\pi_T(r)\{P_{11}(r)-q_wq_h\}$. \qed

\paragraph{Proposition~\ref{prop:ot}.} Every $\gamma_r\in\Gamma_r$ is a joint distribution consistent with the observed margins, so every target compatible with the data has $\PATE_p=\sum_r\pi_T(r)\sum\gamma_r\tau_p$ for some $\gamma_r\in\Gamma_r$, and every such $\gamma$ is compatible. $\Gamma_r$ is a nonempty compact convex polytope and the objective is linear, so its image is the closed interval between the minimum and the maximum, which are attained at vertices. The problems separate across race groups. \qed

\paragraph{Corollary~\ref{cor:frechet}.} The additive terms depend only on the margins. For the interaction, any $P_{11}$ in the Fr\'echet interval is attained by the coupling $\gamma(a,b)=P(d_a,d_b)\,p_w(a\mid d_a)\,p_h(b\mid d_b)$, where $P(\cdot,\cdot)$ is the $2\times2$ table with the given margins and $P_{11}$, and $d_a,d_b$ are the diabetes values of profiles $a,b$; its row and column sums are $p_w(a)$ and $p_h(b)$. No coupling gives $P_{11}$ outside the Fr\'echet interval. For a fixed odds ratio, the $2\times2$ table with given margins is unique and $P_{11}$ solves Plackett's quadratic. \qed

\paragraph{Theorem~\ref{thm:balance}.} On each component $K$, the values $\tau(a,b)$ on the edges determine $(\alpha,\beta)$ up to $\alpha_a+c_K$, $\beta_b-c_K$ (fix one node, propagate along a spanning tree; edges outside the tree add no new freedom). Hence $\Psi$ is determined up to $\sum_Kc_K(w_K-h_K)$, where $w_K=\sum_{a\in K}p_w(a)$ and $h_K=\sum_{b\in K}p_h(b)$. This is zero for all $(c_K)$ if and only if $w_K=h_K$ for every $K$. If some $w_K\ne h_K$, two parameter values that agree on $S$ give different $\Psi$, so $\Psi$ is not identified. \qed

\paragraph{Proposition~\ref{prop:cycle}.} The model restricted to $S$ is linear in $(\alpha,\beta,\eta)$. The parameter $\eta$ is identified if and only if $h|_S$ is not in the space $\mathcal A_S=\{f(a)+g(b)\text{ on }S\}$. A function $\phi$ on the edges lies in $\mathcal A_S$ if and only if its alternating sum around every cycle of $G$ is zero: if $\phi=f+g$, each node's value enters a cycle once with each sign; conversely, if all alternating sums vanish, define $f,g$ on a spanning forest by propagation, and every non-tree edge closes a cycle whose zero sum makes $\phi=f+g$ on that edge. If $\eta$ is not identified, $h=f+g$ on $S$ and the parameters $(\alpha+\eta f,\beta+\eta g,0)$ and $(\alpha,\beta,\eta)$ agree on $S$, so the observable effects fix $\tau_0(a,b)=\alpha_a+\beta_b+\eta\{f(a)+g(b)\}$ on all pairs (by Theorem~\ref{thm:balance}, up to shifts that do not change $\Psi$), and $\Psi(\eta)=\sum\pi_T\tau_0+\eta\sum\pi_T(h-f-g)$. Different $(f,g)$ that represent $h$ on $S$ differ by component shifts, which change $\Delta$ by $\sum_Kc_K(w_K-h_K)=0$. \qed

\paragraph{Corollary~\ref{cor:joint}.} In the index-patient case $h=D_wD_h$ and $f+g=D_w+D_h-1$ on $S$, so $h-f-g=(1-D_w)(1-D_h)$ and $\Delta=\pi_T(00)=\sum_r\pi_T(r)\{1-q_w-q_h+P_{11}(r)\}$. $\PATE(0)$ is additive in the partners and so depends only on the margins. For fixed $\eta$ the target effect is linear in $P_{11}$, and for fixed $P_{11}$ it is linear in $\eta$; a bilinear function on a rectangle attains its extremes at the corners, and every intermediate value is attained by continuity. \qed

\paragraph{Proposition~\ref{prop:shrink}.} By linearity, $E\tilde\tau_j=\lambda_j\tau_j+(1-\lambda_j)\bar\mu$, so $E\sum_j\pi_T(j)\tilde\tau_j-\sum_j\pi_T(j)\tau_j=\sum_j\pi_T(j)(1-\lambda_j)(\bar\mu-\tau_j)$. \qed

\section{Stan program}
The main model is reproduced below; the supplementary code also contains the standard shrink-to-zero comparison and a version with a pooling switch.
\begin{lstlisting}
// Bayesian dyadic multilevel regression and poststratification (dyadic MRP)
// for transporting a couple-randomized trial to a survey-defined population of couples.
//
// Outcome model: bivariate normal model for the wife's and husband's outcomes, fitted to
//   cell-by-arm sufficient statistics. Each partner's treatment effect depends on own traits,
//   partner traits, an own-by-partner diabetes interaction, and couple race/ethnicity.
//   Dyadic partial pooling: the wife's and the husband's coefficient for the same term are
//   pooled toward a shared couple-level mean, with one pooling scale per batch of terms.
// Composition model: Dirichlet model for the joint couple-cell distribution,
//   pi ~ Dirichlet(n_eff * [(1 - omega) * pi_ht + omega * m]), drawn in generated quantities.

data {
  int<lower=1> C;                         // number of couple cells
  int<lower=1> KZ;                        // number of effect-modifier columns
  int<lower=1> KW;                        // number of prognostic columns
  matrix[C, KZ] Zw;                       // modifier design, wife (reference coding)
  matrix[C, KZ] Zh;                       // modifier design, husband
  matrix[C, KW] Ww;                       // prognostic design, wife
  matrix[C, KW] Wh;                       // prognostic design, husband
  int<lower=1> B;                         // number of batches of modifier terms
  array[KZ] int<lower=1, upper=B> batch;  // batch of each modifier column

  // trial: cell-by-arm sufficient statistics
  int<lower=1> G;
  array[G] int<lower=1, upper=C> g_cell;
  vector<lower=0, upper=1>[G] g_arm;
  vector<lower=1>[G] g_n;
  matrix[G, 2] g_mean;                    // (wife, husband) means
  matrix[2, 2] S_tot;                     // pooled within-group scatter matrix
  int<lower=1> N_trial;

  // target survey
  vector<lower=0>[C] dir_par;             // Dirichlet parameters for the composition posterior
  vector<lower=0>[C] pi_ht;               // weighted cell shares (plug-in)
  vector<lower=0>[C] pi_indep;            // plug-in shares, independent pairing within race
  vector<lower=0>[C] pi_trial;            // trial cell shares

  // subgroups for couple-average CATEs
  int<lower=0> NS;
  matrix[C, NS] sub;                      // 0/1 cell membership
}

transformed data {
  matrix[G, KZ] gZw = Zw[g_cell];
  matrix[G, KZ] gZh = Zh[g_cell];
  matrix[G, KW] gWw = Ww[g_cell];
  matrix[G, KW] gWh = Wh[g_cell];
}

parameters {
  vector[2] alpha;            // control-arm intercepts
  vector[KW] beta_w;          // prognostic coefficients
  vector[KW] beta_h;
  vector[2] tau0;             // effect in the reference cell
  vector[KZ] theta_bar;       // shared couple-level modifier coefficients
  vector[KZ] z_w;             // wife deviations (non-centered)
  vector[KZ] z_h;             // husband deviations
  vector<lower=0>[B] sd_dev;  // pooling scale per batch
  vector<lower=0>[2] sigma_y;
  cholesky_factor_corr[2] L_y;
}

transformed parameters {
  vector[KZ] theta_w = theta_bar + sd_dev[batch] .* z_w;
  vector[KZ] theta_h = theta_bar + sd_dev[batch] .* z_h;
}

model {
  alpha ~ normal(0, 10);
  beta_w ~ normal(0, 10);
  beta_h ~ normal(0, 10);
  tau0 ~ normal(0, 10);
  theta_bar ~ normal(0, 5);
  z_w ~ std_normal();
  z_h ~ std_normal();
  sd_dev ~ normal(0, 1.5);
  sigma_y ~ normal(0, 20);
  L_y ~ lkj_corr_cholesky(2);
  {
    matrix[G, 2] R;
    R[, 1] = g_mean[, 1] - (alpha[1] + gWw * beta_w) - g_arm .* (tau0[1] + gZw * theta_w);
    R[, 2] = g_mean[, 2] - (alpha[2] + gWh * beta_h) - g_arm .* (tau0[2] + gZh * theta_h);
    matrix[2, 2] L_S = diag_pre_multiply(sigma_y, L_y);
    matrix[2, 2] Sinv = chol2inv(L_S);
    real quad = sum(g_n .* rows_dot_product(R * Sinv, R));
    target += -N_trial * sum(log(diagonal(L_S))) - 0.5 * trace(Sinv * S_tot) - 0.5 * quad;
  }
}

generated quantities {
  vector[3] pate_full;     // wife, husband, couple average
  vector[3] pate_fixed;
  vector[3] pate_indep;
  vector[3] sate_model;
  vector[NS] tau_sub;      // couple-average CATEs in target subgroups
  real rho_y = multiply_lower_tri_self_transpose(L_y)[1, 2];
  {
    vector[C] tw = tau0[1] + Zw * theta_w;
    vector[C] th = tau0[2] + Zh * theta_h;
    vector[C] tc = 0.5 * (tw + th);
    vector[C] pm = dirichlet_rng(dir_par);
    pate_full  = [dot_product(pm, tw), dot_product(pm, th), dot_product(pm, tc)]';
    pate_fixed = [dot_product(pi_ht, tw), dot_product(pi_ht, th), dot_product(pi_ht, tc)]';
    pate_indep = [dot_product(pi_indep, tw), dot_product(pi_indep, th), dot_product(pi_indep, tc)]';
    sate_model = [dot_product(pi_trial, tw), dot_product(pi_trial, th), dot_product(pi_trial, tc)]';
    for (s in 1:NS) tau_sub[s] = dot_product(pm .* col(sub, s), tc) / dot_product(pm, col(sub, s));
  }
}
\end{lstlisting}

\end{document}